\documentstyle[11pt,epsfig]{article}

\newcommand{\reals}{\mbox{R}}

\newcommand{\integers}{\mbox{Z}}

\newcommand{\beq}{\begin{equation}}
\newcommand{\eeq}{\end{equation}}
\newcommand{\beqa}{\begin{eqnarray}}
\newcommand{\eeqa}{\end{eqnarray}}
\newcommand{\bi}{\begin{itemize}}
\newcommand{\ei}{\end{itemize}}
\newcommand{\di}{\displaystyle}
\newcommand{\tr}{\mbox{tr}}
\newcommand{\pexp}{\mbox{Pexp}}
\newcommand{\fot}{\frac{1}{2}}

\def\fon{\frac{1}{N}}
\def\a{\alpha}
\def\b{\beta}
\def\c{\gamma}
\def\d{\delta}
\def\e{\epsilon}
\def\l{\lambda}

\def\pc{\psi_\times}
\def\pp{\psi_+}
\def\pmi{\psi_-}
\def\pn{\psi_\asymp}

\def\pupdown{\psi_{\stackrel{\vee}{\scriptscriptstyle\wedge}}}
\def\ptp{\psi_{\propto+}}
\def\ptm{\psi_{\propto-}}
\def\intA{\int \!DA\,}
\def\D{\tilde D}
\def\DD{\tilde D'}

\begin{document}

\begin{flushright}
MPI-Ph/93-107
\\
hep-th/mmyynnn
\\
December 1993
\end{flushright}
\vfill

\begin{center}{\Large\bf 
On a geometric derivation of Witten's identity for Chern-Simons theory}
\end{center}
\vfill

\begin{center}
\large Bernd Br\"ugmann
\end{center}
\smallskip

\begin{center}
{
\em Max-Planck-Institut f\"ur Physik, 80805 M\"unchen, Germany
\\
bruegman@iws170.mppmu.mpg.de
} 
\end{center}
\vfill

\begin{center}
\large\bf Abstract
\end{center}
\medskip

\noindent
We present a formal but simple calculational scheme to relate the
expectation value of Wilson loops in Chern-Simons theory to the Jones
polynomial. We consider the exponential of the generator of homotopy
transformations which produces the finite loop deformations that define the
crossing change formulas of knot polynomials.  Applying this operator to
the expectation value of Wilson loops for an unspecified measure we find a
set of conditions on the measure and the regularization such that the Jones
polynomial is obtained.

\vspace*{\fill}
\newpage

\section{Introduction}

Knot theory has recently played an important role in physics. In knot
theory we study the equivalence classes of loops under diffeomorphisms
connected to the identity, hence knot theory can be expected to appear
naturally in the context of a diffeomorphism invariant theory that deals
with loops. Loops are of interest in any gauge theory because the trace of
the parallel transport operator around a loop is a gauge invariant
variable, the so-called Wilson loop. Prime examples for a diffeomorphism
and gauge invariant theory are topological field theory \cite{Wi89} in three
dimensions and canonical quantum gravity \cite{RoSm88} in four dimensions,
and in both cases knot theory figures prominently.

The arguably most intriguing connection between knot theory and field
theory is encapsulated in Witten's identity \cite{Wi89} that expresses the
Jones polynomial $J_\c(q)$ as the vacuum expectation value of a Wilson loop
$W_\c[A]$ in a Chern-Simons theory,
\beq
	\langle W_\c[A] \rangle_{k,N} = \a^{-w(\c)} J_\c(q).
\label{wittenid}
\eeq
Let us define both sides of this equation.

We consider Chern-Simons theory for an $SU(N)$ connection $A_a(x) =
A_a^i(x)T^i$ on a three manifold $\Sigma$, for which we choose $\Sigma =
S^3$. Here the $T^i$ are the generators of the fundamental
representation of $SU(N)$ normalized according to $\tr T^i T^j = \fot
\d^{ij}$, and
\beqa
	\langle W_\c[A] \rangle_{k,N} &=& 
	\intA \exp ( \frac{ik}{4\pi} S_{CS}[A] ) W_\c[A] ,
\\
	S_{CS}[A] &=& \int_\Sigma \!d^3x\, \e^{abc} \tr (A_a\partial_bA_c
	+ \frac{2}{3} A_aA_bA_c),
\eeqa
where $k\in\integers$ is the coupling constant. The Chern-Simons action has
the characteristic property that
\beq
	\frac{\d}{\d A_a^i(x)} S_{CS}[A] = \fot \e^{abc} F_{bc}^i(x),
\label{dsda}
\eeq
where $F_{bc}^i$ is the curvature of $A_a^i$.

We define a path to be a continuous, piecewise smooth map
$\c:[s,t]\rightarrow\Sigma$ with non-vanishing tangent $\dot\c^a(s)$, and a
loop is defined as a closed path, $\c:[0,1]\rightarrow\Sigma$ with
$\c(0)=\c(1)$.  Loops are allowed to have intersections (here we consider
only the case of double points). A Wilson loop is defined by
\beqa
	W_\c[A] &=& \tr U_\c(0,1) \equiv \tr U_\c,
\\
	U_\c(s,t) &=& \pexp \int_0^1 \!ds\, \dot\c^a(s) A_a(\c(s)),
\eeqa
where $U_\c$ is the parallel transport operator along the loop $\c$, and P
denotes path-ordering. The Wilson loops satisfy the following
differential equations:
\beqa
	\frac{\d}{\d A_a^i(x)} \tr U_\c &=&
	\int \!ds\, \dot\c^b(s) \d^3(x, \c(s)) \tr U_\c(s) F_{ab}(\c(s)),
\label{dwda}
\\
	\frac{\d}{\d \c^a(s)} \tr U_\c &=&
	\dot\c^b(s) \tr U_\c(s) F_{ab}(\c(s)),
\label{dwdl}
\eeqa
where $U_\c(s)$ is the parallel transport once around the loop from $s$ to
$s$.

On the right hand side of Witten's identity (\ref{wittenid}) we have two
constants that are determined by the parameters of the Chern-Simons theory,
\beq
	q = \exp\frac{i\pi}{N+k}, \quad \a = q^{N-\frac{1}{N}}.
\label{qa}
\eeq 
The whole expression is directly related to the Kauffman bracket
\cite{Ka91}, and it is {\em not} a diffeomorphism invariant functional of
loops.  The Kauffman bracket is defined not for knots in three dimensions
but only for knot diagrams. One usually defines invariants of knot diagrams
which arise from links, i.e.\ from multiloops $\c:S^1\times\ldots\times
S^1\rightarrow\Sigma$.

\begin{figure}
\centerline{\mbox{\epsfig{figure=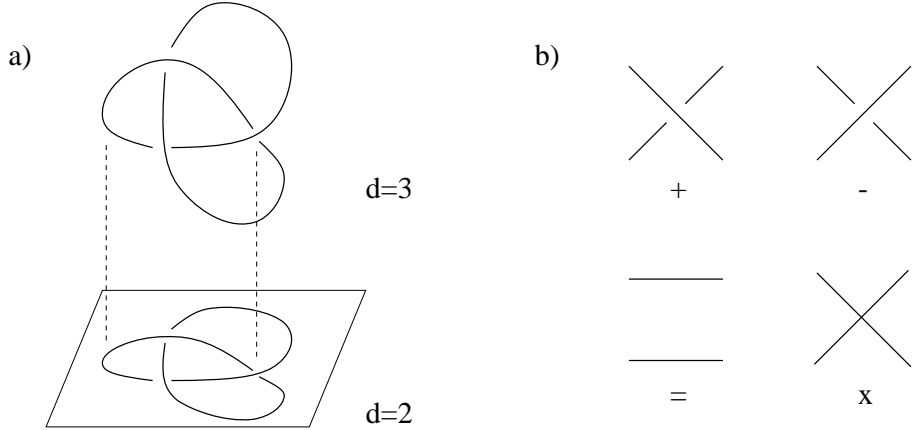,width=120mm}}}
\caption{a) Projecting a knot onto a knot diagram; b) the four crossings in
a knot diagram.}
\label{f1}
\end{figure}

A knot diagram is a regular projection of a loop into a plane together with
labels that distinguish intersections that arise from overcrossings ($+$)
or undercrossings ($-$) (figure \ref{f1}).  We also introduce labels for
the genuine intersection of two lines ($\times$) and for two
non-intersecting lines ($\asymp$). The unknot is denoted by ($\circ$).
When discussing framing we denote a positive twist by ($\propto+$), a
negative twist by ($\propto-$), the intermediate intersection by
($\propto$), and the untwisted line by ($\wr$).

The winding number $w(\c)$ is equal to the sum over all crossings in a knot
diagram counting $+1$ for overcrossings and $-1$ for undercrossings, and it
is only a regular isotopy invariant. While $w(\c)$ does not change under
small deformations, it does depend on the projection which may introduce an
arbitrary number of crossings into the knot diagram. This type of
projection dependence is also called framing dependence.

The knot polynomial $J_\c(q)$ is an ambient isotopy invariant and hence does
not depend on the projection.  It is a Laurent polynomial in one complex
variable $q$ defined by
\beqa
	J_\c &=& J_{\c'} \quad \mbox{if $\c\sim\c'$ in three space},
\label{diffic}
\\
	J_\circ &=& \frac{q^N - q^{-N}}{q - q^{-1}},
\label{normal}
\\
	q^N J_+ - q^{-N} J_- &=& (q-q^{-1}) J_\asymp.
\label{skein}
\eeqa
The original Jones polynomial \cite{Jo85} is obtained for $N=2$.  The
crossing change formula (or skein relation) (\ref{skein}) allows us to
recursively reduce any knot diagram to the polynomial of the unknot, which
is defined by (\ref{normal}). As usual, the crossing change formula relates
polynomials for projected links that differ only in one four-valent
diagram, where the knot assumes one of the elementary crossings ($+$),
($-$), or ($\asymp$).

The right hand side of Witten's identity, 
\beq
	K_\c(q) = \a^{-w(\c)} J_\c(q), 
\eeq
is therefore the regular isotopy invariant defined by
\beqa
	K_\c &=& K_{\c'} \quad \mbox{if $\c\sim\c'$ under regular isotopy},
\label{invk}
\\
	K_\circ &=& \frac{q^N - q^{-N}}{q - q^{-1}},
\label{normk}
\\ 
	q^\fon K_+ - q^{-\fon} K_- &=& (q-q^{-1}) K_\asymp,
\label{skeink}
\\
	K_{\propto\pm} &=& \a^{\pm1} K_{\wr},
\label{twistk}
\eeqa
since $w_+ - 1 = w_\asymp = w_- + 1$ and $J_{\propto\pm}=J_\wr$.

We have defined all parts of Witten's identity except for how the path
integral is to be computed. In particular, we did not specify the measure
$DA$. This is, of course, the crucial input that determines the outcome of
the calculation. Let us mention four approaches towards defining the path
integral. 
\begin{enumerate}

\item
First of all, there is Witten's original paper \cite{Wi89}, in
which he proves (\ref{wittenid}) by arguments from conformal field theory
for the 2+1 decomposition of Chern-Simons theory without explicitly
defining the measure $DA$. Witten's beautiful argument is convincing,
although some of the mathematical details still have to be worked out
(e.g.\ \cite{Sc93}) (which applies equally well or even more so to
(2.)--(4.)). 

\item
A quite unrelated approach is that of standard perturbation theory, in
which a measure and a gauge fixing is defined explicitly (e.g.
\cite{AlLaRa90,Bar90,GuMaMi90,AxSi92}). A proof of Witten's identity to all
orders has been presented in \cite{AxSi93}. A recurrent problem in the
perturbation theory of non-abelian Chern-Simons theory is to reproduce the
shift of $k$ to $k+N$ in (\ref{qa}) \cite{AlLa90,Gu93}.

\item
At the same time it was realized that geometric deformations of the loop
inside the path integral are directly related to perturbative results
\cite{Sm89,Aw90,CoGuMaMi90,Ka93}. At least to first order in $1/k$, 
deformations of the loop lead to the same results as perturbation theory,
and Witten's identity is satisfied to this order. Into this category falls
the attempt to solve the analog of the Makeenko-Migdal loop equation known
from Yang-Mills theory and thereby to obtain Witten's identity to all
orders in $1/k$ \cite{Aw90}.

\item
An order by order analysis in perturbation theory shows that the
coefficients in a $1/k$ expansion define particular knot invariants
\cite{GuMaMi90,Bar90}. These knot invariants are closely related to
so-called Vassiliev invariants \cite{Va90,Bae92,Bar92,Ka93}. Recently, it
has been shown (see \cite{Bi93}) that knot polynomials like the Jones
polynomial arise as power series with Vassiliev invariants as
coefficients. Implicit in this construction seems to be a proof that the
perturbation series sums up to the Kauffman bracket.

\end{enumerate}
Each of these approaches has its own merit, since the different techniques
have lead to different, interesting insights into the relation between
Chern-Simons theory and knot theory. 

The purpose of this paper is to present a formal calculational scheme based
on loop deformations that allows to derive Witten's identity to all orders
in $1/k$. As such it can be viewed as an extension of the third, the
geometric approach.  The hope is that since $K_\c(q)$ is a rather simple
function of loops, a simple, intuitive argument may arise by focusing on
the loop dependence of $\langle W_\c[A] \rangle$. This is indeed the case.

Let us give a brief outline of our method. The starting point for
our construction is the observation that the generator of homotopy
transformations, $\D$, applied to
\beq
	\psi_\c =\langle W_\c[A] \rangle,
\eeq
reproduces the
perturbation expansion to linear order in $1/k$.  We define the exponential
of $\D$, which generates finite deformations of parts of a loop along a
vector $v$. Such an operator can, for example, lift one line of a true
intersection thereby transforming ($\times$) to ($+$), i.e.
for a suitable choice of $v$ we have that
\beq
	e^{\D(v)} \pc = \pp.
\label{l1}
\eeq
Under natural assumptions about the path integral and the regularization
procedure we derive from the calculation of 
$\langle \D(v) W_\c[A] \rangle$ that
\beq
	e^{\D(v)} \pc = a \pc + b \pn
\label{l2}
\eeq
for some coefficients $a$ and $b$. Finally, we show that there exists an
essentially unique regularization such that from
(\ref{l1}) and (\ref{l2}) follow precisely the skein relations that define
$K_\c(q)$ for $q = \exp\frac{i\pi}{k}$. 

Our method must definitely be called formal, since we neither give an
indirect definition of the path integral of Chern-Simons theory as in (1.),
nor an explicit definition as in (2.). We assume that there exists an
otherwise unspecified measure $DA$ such that $\langle W_\c[A] \rangle $
exists. The goal is {\em to find a minimal set of assumptions} about the
measure and regularization such that Witten's identity is reproduced. 

This approach also differs from (4.) in that we do not analyze knot
invariants at each order but only the summed series. A common
feature is, however, that {\em loops with intersections} play a natural
role. Naively, this should be expected simply because the classes of knots
with intersections `separate' the classes of non-intersecting
knots. However, the great power of such a point of view in knot theory was
only realized very recently (considering the long history of knot theory)
in Vassiliev's work \cite{Va90,Bi93}. The definition of the Vassiliev
invariants makes crucial use of intersecting knots, and there is a
conjecture that for the first time a complete set of knot invariants may be
obtained from such invariants (see \cite{Bi93}).

In physics, a motivation to study knot invariants for loops with
intersections arises in the study of the loop representation of canonical
quantum gravity in 3+1 dimensions \cite{RoSm88,RoSm90}. It is interesting
to note that (to the knowledge of the author) an extension of the braid group
to intersections appeared for the first time in \cite{Sm88}, and a
definition of the Jones polynomial based on intersecting braids in
\cite{Ga92}. The states in the loop representation are functionals of 
loops, $\psi[\eta]$, and the space of solutions to the spatial
diffeomorphism constraint is the space of knot invariants. Loop functionals
of non-intersecting loops, however, are annihilated by the determinant of
the metric (which in particular implies that they are solutions for
arbitrary cosmological constant
\cite{BrPu91}). Non-degenerate solutions to both the diffeomorphism
constraint and the Wheeler-DeWitt equation of canonical quantum gravity
arise for loops with a generic triple self-intersection
\cite{BrGaPu92a}. Furthermore, these solutions are formally related to the
Chern-Simons action \cite{BrGaPu92b}, which gives rise to a solution to the
constraints \cite{Ko90} in what is known as the connection representation
of quantum gravity \cite{As91}. In this context an extension of Witten's
identity to intersecting loops becomes necessary, and has been given up to
order $1/k$ in \cite{BrGaPu92b}. As a by-product of our method we find that
Witten's identity holds (to all orders) for the Jones polynomial of
intersecting loops defined in \cite{Ga92}.

We will proceed as follows. In section 2, we recall how first order loop
deformations of $\langle W_\c[A] \rangle$ can be computed and
interpreted. In section 3, we introduce the generator of homotopy
transformations. In section 4, we present the formal calculation that leads
to Witten's identity. We conclude with a discussion in section 5.

\section{Formal first order calculation}

Let us reproduce the type of calculation that shows how loop deformations
in the Chern-Simons expectation value of Wilson loops lead to a first order
result that is consistent with the skein relations that define the Jones
polynomial. The approach taken here is most closely related to
\cite{CoGuMaMi90,BrGaPu92b}. 
First we compute the functional derivative with respect to the loop of the
expectation value of a single Wilson loop $W_\eta$:
\beqa
	&&	
\frac{\d}{\d\eta^a(s)} 
\langle W_\eta[A] \rangle
\nonumber
\\
	&=&  
\intA 
\exp(\frac{ik}{4\pi} S_{CS}[A])
\frac{\d}{\d\eta^a(s)} 
\tr U_\eta
\label{exchange}
\\
	&=&
\dot\eta^b(s)
\intA 
\exp(\frac{ik}{4\pi} S_{CS}[A])
F^j_{ab}(\eta(s)) 
\tr U_\eta(s) T^j
\label{diff1}
\\
	&=&
- \frac{4\pi i}{k} 
\e_{abc}
\dot\eta^b(s)
\intA
\left(
\frac{\d}{\d A_c^j}
\exp(\frac{ik}{4\pi} S_{CS}[A])
\right) 
\tr U_\eta(s) T^j
\label{csprop}
\\
	&=&
\frac{4\pi i}{k} 
\e_{abc}
\dot\eta^b(s)
\intA
\exp(\frac{ik}{4\pi} S_{CS}[A])
\frac{\d}{\d A_c^j}
\tr U_\eta(s) T^j
\label{partial}
\\
	&=&
\frac{4\pi i}{k} 
\e_{abc}
\dot\eta^b(s)
\int\!dt\, \dot\eta^c(t) \d^3(\eta(s),\eta(t))
\intA 
\exp(\frac{ik}{4\pi} S_{CS}[A])
\tr U_\eta(s,t)T^j U_\eta(t,s) T^j 
\label{diff2}
\nonumber
\\ && \\
	&=&
\frac{2\pi i}{k}
\e_{abc}
\dot\eta^b(s)
\int\!dt\, \dot\eta^c(t) \d^3(\eta(s),\eta(t))
(\langle \tr U_\eta(s,t) \tr U_\eta(t,s) \rangle 
- \frac{1}{N}
\langle \tr U_\eta \rangle ) .
\label{lova}
\eeqa
Here we have used (\ref{dwdl},\ref{dsda},\ref{dwda}) in
(\ref{diff1},\ref{csprop},\ref{diff2}), respectively. 
The last step is specific to $SU(N)$, for which 
\beq
	T^j_{ab}T^j_{cd} = 
	\fot (\d_{ad}\d_{bc} - \frac{1}{N}\d_{ab}\d_{cd}).
\eeq
The pretty result is that {\em the loop variation of $\langle W_\eta
\rangle$ can be again expressed in terms of expectation values of Wilson
loops}. This happens precisely because the Chern-Simons action has the
property (\ref{dsda}), which is not true for a generic weight in the path
integral.

Since we want to keep track of what is rigorous and what formal, let us
emphasize that two crucial assumptions have been made in (\ref{exchange})
and (\ref{partial}):
\begin{itemize}
\item[(A1)]
The limits of differentiation and integration commute,
\beq
	\frac{\d}{\d\eta^a(s)} 
        \langle W_\eta[A] \rangle
	=
	\langle \frac{\d}{\d\eta^a(s)} W_\eta[A] \rangle.
\eeq

\item[(A2)]
There are no contributions from boundary terms or the measure in 
partial integrations of $\d/\d A_c^j$,
\beq
	\intA 
	\frac{\d}{\d A_c^j}
	\left(
	\exp(\frac{ik}{4\pi} S_{CS}[A])
	\tr U_\eta(s) T^j
	\right) 
	= 0.
\eeq
\end{itemize}
As we will explain below, both assumptions are wrong in general.

The result of the loop deformation can be given a rough interpretation as
follows, which we will make more precise in the next section.  Consider a
loop $\eta$ with a transverse intersection where two (but not more) lines
meet, i.e. for some $s$ and $t$, $s\neq t$, we have $\eta(s) = \eta(t)$ and
$\dot\eta(s) \neq \dot\eta(t)$.  Let us assume that the effect of the
operator $v^a\d/\d\eta^a(s)$ is to lift one of the lines that run through
the intersection along the direction $v^a$ such that by definition
\beq
	v^a \frac{\d}{\d\eta^a(s)} \pc = \pp - \pc .
\eeq
Focusing on a single intersection, and assuming that at the point where
the two loops in
$	\pupdown \equiv 
	\langle \tr U_\eta(s,t) \tr U_\eta(t,s) \rangle 
$
touch we can smooth out the corners such that $\pn = \pupdown$,
we can write (\ref{lova}) as
\beq
 		v^a \frac{\d}{\d\eta^a(s)} \pc = 
		\frac{2\pi i V}{k}  (\pn - \fon \pc).
\label{primd}
\eeq
One now argues that it is a matter of proper regularization to assign a
finite value to the `volume element' 
$
V = 
\int\!dt\, \e_{abc} v^a
\dot\eta^b(s)
\dot\eta^c(t) \d^3(\eta(s),\eta(t))
$, say $V = 1/2$. 

Hence loop deformations at an intersection in the
directions $v^a$ and $-v^a$ lead to 
\beqa
	\pp = (1-\frac{\pi i}{kN} )\pc + \frac{\pi i}{k} \pn ,
\\
	\pmi = (1+\frac{\pi i}{kN} )\pc - \frac{\pi i}{k} \pn.
\eeqa
In order to obtain a crossing change formula for knots without
intersections, we eliminate $\pc$ and obtain
\beq
(1+\frac{\pi i}{kN} )\pp - (1-\frac{\pi i}{kN} )\pmi =  \frac{2\pi i}{k} \pn.
\eeq
In fact, this crossing change formula is to linear order in $1/k$ exactly
the crossing change formula for $\psi_\c = K_\c$, (\ref{skeink}),
\beq
	q^\fon \pp - q^{-\fon} \pmi = (q-q^{-1}) \pn.
\eeq
The framing dependence of $\psi_\c$ can be derived by a similar argument
(see section 3), and is found to agree to linear order with that of $K_\c$,
(\ref{twistk}). 

The conclusion is that, given certain assumptions, (i) we can compute the
effect of loop deformations on $\langle W_\c \rangle$ in `closed' form,
(ii) we can interpret the result as a skein relation for $\psi_\c = \langle
W_\c \rangle$, and (iii) the coefficients in the skein relations are to
linear order in $1/k$ the ones that appear in Witten's identity. 

As already explained in the introduction, our goal is to explore under what
conditions higher order loop deformations lead to higher order corrections
in $1/k$ that sum to the correct coefficients.  To this end let us now give
a more precise definition of the loop deformations that take intersections
apart.

\section{Generating homotopy transformations}

Consider the following two generators of infinitesimal transformations of
loops:
\beqa 
	D(w) \psi[\eta] &=& \int\!ds\, w^a(\eta(s)) \frac{\d}{\d\eta^a(s)}
		\psi[\eta],
\\
	\D(v) \psi[\eta] &=& \int\!ds\, v^a(s) \frac{\d}{\d\eta^a(s)}
		\psi[\eta].
\eeqa
The operator $D(w)$ is the natural generator of diffeomorphisms on the
space of loop functionals. Each point $\eta(s)$ of the loop is displaced an
infinitesimal amount along a vector field $w^a\in T\Sigma$, and $D(w)$
satisfies the algebra of diffeomorphisms, $[D(w),D(w')] = D({\cal L}_w(w'))$.
Obviously, intersections of loops are invariant under the action of $D(w)$,
since, independently of the parameters $s$ and $t$, 
$\eta(s)=\eta(t)$ is moved as one single point.

On the other hand, $\D(v)$ generates a more general type of transformation,
since now $v^a$ assigns a vector to each parameter value of the loop (as
opposed to a vector to each point in the manifold). This is exactly what we
need to deform loops from intersecting to non-intersecting. For example, we
can choose $v^a$ to vanish everywhere along the loop except in the
neighborhood of {\em one} of the legs of an intersection such that
$\eta(s)$ is moved but $\eta(t)$ remains in place.

$\D(v)$ is the generator of homotopy transformation in loop space $L$.  To
be more precise, two loops are homotopic if they can be continuously
deformed into each other. This will be the case for continuous $v^a$, which
also can remove corners in loops. Smooth $v^a$ lead to smooth deformations.
While the orbits of $D(w)$ are curves on the three manifold, the orbits of
$\D(v)$ are curves in $L$. A curve in $L$ is defined by a one
parameter family of loops, $\eta_u$, and
\beq
	\frac{d}{du} \psi[\eta_u]
	= \int\!ds\, \frac{\partial\dot\eta_u^a(s)}{\partial u} 
	\frac{\d}{\d\eta_u^a(s)} \psi[\eta_u] 
	= \D(v) \psi[\eta_u], 
\eeq
for $\di v^a(s) = \frac{\partial\dot\eta^a_u(s)}{\partial u}$.

\begin{figure}
\centerline{\mbox{\epsfig{figure=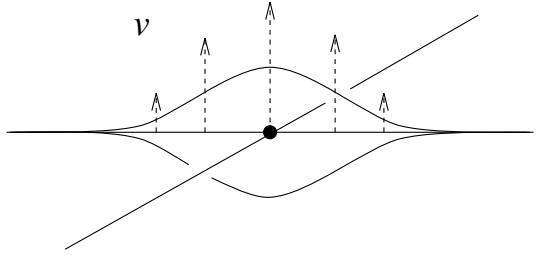,width=70mm}}}
\caption{A family of loops that defines a crossing change.}
\label{f2}
\end{figure}

\begin{figure}
\centerline{\mbox{\epsfig{figure=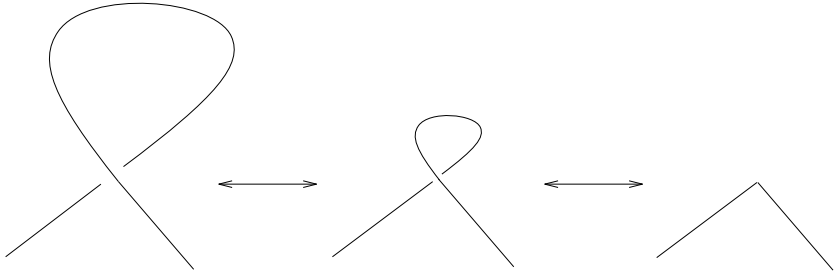,width=110mm}}}
\caption{A family of loops that defines the removal of a twist by changing
a crossing to an intersection and simultaneously shrinking the twisted part
of the loop to a point.}
\label{f3}
\end{figure}

Our plan is to use $\D(v)$ as a concretization of the loop deformations of
the previous section. For example, we can define a family of curves that
interpolate between an overcrossing and an undercrossing (with respect to
some preferred direction, figure \ref{f2}). The corresponding generator we
denote by just $\D$, and from (\ref{primd}),
\beq
	\D \pc = \frac{\pi i}{k} (\pn - \fon \pc).
\label{dp}
\eeq 

With a small modification we can also define a family of loops that stands
for introducing a twist ($\propto\pm$) into a line ($\wr$) (figure
\ref{f3}).  We obtain the correct framing factor if (in inverted order) we
first make the transition from crossing ($\propto+$) to intersection
($\propto$), then shrink the loop to a corner ($<$), evaluate the path
integral, and smooth the corner to ($\wr$). Notice that this is a different
procedure then the one that defines the transition from $\psi_\propto$ to
$\psi_<$ (for which the correct result is (\ref{twistfactor})).
From (\ref{lova}) we derive for the corresponding generator $\DD$ that
\beq
	\DD \psi_< = \frac{\pi i}{k} (N -\fon) \psi_<.
\label{ddp}
\eeq

It is worth emphasizing that both generators $\D$ and $\DD$ refer to an
essentially planar representation of a localized crossing. The family of
loops for $\D$ may only introduce a single crossing, and in this case the
crossing can be made arbitrarily flat in any coordinate system. For
the family of loops defining $\DD$ we have in addition to require that the
separation at the crossing and the spread in $v^a(s)$ is much smaller then
the characteristic radius of the twist.

In the remainder of this section we want to comment on the regularization
of the volume element $V$ and on the exponentiation of $\D$.

\subsection{Regularization of the volume element}

The volume element for $\D(v)$ is
\beq
	V = 
\int\!ds \int\!dt\, \e_{abc} v^a(s)
\dot\eta^b(s)
\dot\eta^c(t) \d^3(\eta(s),\eta(t)).
\eeq
There is at least a $\d(0)$ singularity since the three dimensional delta
distribution depends only on two parameters. 

Suppose we ignore the singularity for the moment.  If we set $v^a(s) =
v^a(\eta(s))$, then $V$ is the volume element for a deformation generated
by a diffeomorphism $D(v)$. Since in this case $v^a(s)=v^a(t)$, the
antisymetrization in the tangent vectors implies that $V=0$ and therefore
that $\langle W_\eta[A] \rangle$ is invariant under diffeomorphisms.  We
therefore expect that it is the regularization that introduces the
well-known framing dependence of $\langle W_\eta[A] \rangle$.

There are two classes of singularities in $V$ corresponding to the two
classes of zeroes of $\eta(s) - \eta(t)$ (in coordinates). For ranges of
$s$ and $t$ without intersections, $\eta(s)-\eta(t)=0$ implies $s=t$. A
standard regularization in this case is to replace one of the two
integrations along $\eta$ by an integration around the framed loop
$\eta^f(s)$, which is obtained from $\eta(s)$ by displacing $\eta(s)$ in
some direction not parallel to the loop without introducing intersections.
The different possibilities of framing are labeled by the linking number of
the framed and unframed loop, which is an integer. Since by definition
$\eta(s)$ and $\eta^f(t)$ do not intersect, we conclude that $V=0$, and
therefore that $\langle W_\eta[A] \rangle$ is invariant under
diffeomorphisms when regularized by framing.

In the case that $\eta(s)$ has a self-intersection at $s_0\neq t_0$, we
have $V=0$ as before for a framing such that $\eta(s) \neq \eta^f(t)
\,\,\forall s, t $. However, we are interested in the situation where
$\eta(s)$ is replaced by a family of curves $\eta_u(s)$ that describe a
crossing change. In the presence of a framing, the family of curves
$\eta_u(s)$ has to cross $\eta^f(t)$, and therefore there does occur a
singularity for some $u$. This singularity is independent of the
considerations that lead to the framing, and we prefer to discuss it in the
limit in which the framing has been removed. (Another convenient way to
combine framing with crossing change is to specialize the framing such that
near a self-intersection the framed loop has to maintain an intersection
with the unframed loop, e.g. for a loop which is planar near an
intersection the direction of displacement of the framing must be co-planar
with the tangent vectors at the intersection.)

So let us consider the case where $\eta_u$ describes a crossing change and
the framing has been removed.
The deformations we consider are along $v^a(s) =
\frac{\partial\dot\eta_u^a(s)}{\partial u}$ and are localized on one leg
close to a crossing. Referring to a particular direction, let $\eta_{-1}$
run below, $\eta_1$ above, and $\eta_0$ through the intersection at
$\eta_0(s_0) = \eta_0(t_0)$. These three cases correspond to $\pmi$, $\pp$,
and $\pc$, respectively (compare figure \ref{f2}).

In order to obtain a finite answer we define $\psi[\eta]$ for the
transition from $\pc$ to $\pp$ by a `smearing' over the strip defined by
the family $\eta_u$,
\beq
	\psi[\eta] =  \int_{0}^{1} \!du\, \psi[\eta_u].
\label{smearing}
\eeq
Then
\beq
	V =  \int_{0}^{1} \!du
\int\!ds \int\!dt\, \e_{abc} v^a(s)
\dot\eta_u^b(s)
\dot\eta_u^c(t) \d^3(\eta_u(s),\eta_u(t)).
\label{V}
\eeq
The delta function can be easily removed if 
$\e_{abc} v^a(s) \dot\eta_u^b(s) \dot\eta_u^c(t) \neq 0 $, since then
\beq
	\d^3(\eta_u(s),\eta_u(t)) = 
	\frac{1}
	{| \e_{abc} v^a(s_0) 
	\dot\eta_u^b(s_0)
	\dot\eta_u^c(t_0) |}
	\d(s-s_0) \d(t-t_0) \d(u-u_0),
\eeq
where for our choice of $v^a$ we obtain only one term representing the sum
over all zeros of $\eta_u(s) - \eta_u(t)$.  

By construction of the crossing change and by the assumption that the
intersection is transverse, the tangent vectors and $v^a$ are indeed not
co-planar. Therefore the final result for the volume element when
regulated by strips is in the case of an intersection 
\beq
\mbox{$V= \fot$ for $\pp$,} \quad
\mbox{$V = -\fot$ for $\pmi$},
\eeq
as required for the first order argument of the preceding section.
(The factor $\fot$ is due to $\int_0^1\!dx\,\d(x)=\fot$.)

Since for the exponentiation of $\D$ we have to compute $\D^n$, we absorb
the smearing (\ref{smearing}) into the definition of $\D$, i.e. each time
$\D$ acts we perform an extra integration over $u$.


\subsection{Exponentiation}

The generator of diffeomorphisms can be exponentiated to give finite
diffeomorphisms connected to the identity. The analog is true for the
generator of homotopy transformations. One way to see this is to notice
that the space of loops inherits a natural differential structure as a
space of mappings between two differentiable manifolds. The operator we
would like to consider is
\beq
	\exp (\D) \psi[\eta] \equiv \exp (\frac{d}{du}) \psi[\eta_u], 
\eeq
where on the left hand side $\eta$ stands for a family of loops which
defines the displacement in $\D$. Naturally, $\exp (\D) \psi[\eta]$ exists
only if $\psi[\eta]$ satisfies an appropriate differentiability condition.
  
The problem is that $\psi[\eta] = \langle W_\eta[A] \rangle$ does not
satisfy the necessary condition. Consider any function
$f:\reals\rightarrow\reals$. Then
\beq
	\exp (c \frac{d}{du}) f(u) := \sum_{n=0}^\infty 
	\frac{c^n}{n!} (\frac{d}{du})^n f(u)
	= f(u+c)
\label{expf}
\eeq
only if $f$ is in fact analytic. However, 
we have just shown that for the regularization involving a framing,
that we want to use, the expectation value of a Wilson loop is a certain
step function near an intersection,
\beq
	f(u) = \langle W_{\eta_u} \rangle = 
	\left\{ 
	\begin{array}{ll}
	\pp & \mbox{if $u>0$} \\
	\pc & \mbox{if $u=0$} \quad , \\
	\pmi & \mbox{if $u<0$} 
	\end{array}
	\right.
\eeq  
where $\pp$, $\pc$, and $\pmi$ are constants independent of $u$.

In other words, what we would like to define is not an operator on $L$, but
on $L$ modulo diffeomorphisms. This is a non-trivial task, since the
quotient space does not inherit a natural differential structure. A similar
problem occurs in the construction of loop representations \cite{BrPu93},
since there the loop states satisfy identities deriving from identities
among the Wilson loops, which are reparametrization invariance and the
Mandelstam identities. In this context a rigorous mathematical framework
for differential calculus is being developed \cite{Ta93}, which involves an
extension to distributional functionals of loops. It is still unclear
whether it is possible to treat diffeomorphism invariance in this
framework.

A natural point of view is that, given that $\langle W_{\eta_u} \rangle$ is
a step function in $u$, we should have expected that $d/du$ leads to a
delta function in $u$. This is precisely what we obtained in section 3.1,
but there we regulated away the $\d(u)$ by an integration over $u$,
(\ref{smearing}). This regularization is performed so that we can identify
the first order loop variations with the finite terms in the first order
expansion in $1/k$. We will return to this issue in section 5.

These observations are, in fact, based on assumption (A1). In order to be
able to perform the calculation (\ref{exchange}--\ref{lova}) of the loop
variations, we have to take the loop variations $\d/\d \eta^a(s)$ inside
the path integral.  In the process, we break diffeomorphism invariance.  If
we had independent knowledge of the step function behaviour of $f(u) =
\langle W_{\eta_u} \rangle$, then we could work completely on the level of
distributional derivatives of $f(u)$.

The approach we take here is to motivate assumption (A1) by making an
additional assumption about the regularization of 
$\langle W_{\eta_u}\rangle$. 
The problem is not to define $\frac{d}{du}\langle W_{\eta_u}\rangle$. 
Notice that any step function can be obtained as the limit of a sequence of
analytic functions. To give a concrete example,
suppose $g(u)$ is a step function that jumps from $a$ to $b$ at
$u=0$, and $g(0)$ is finite, e.g. $g(0) = (a+b)/2$. Then $g(u)$ is the
limiting case, $\lim_{\e\rightarrow0} g_\e(u) = g(u)$, of e.g.
\beq
	g_\e(u) = a + \frac{b-a}{\e\sqrt{\pi}} \int_{-\infty}^{u}\!du \, 
	e^{-x^2/\e^2}.
\eeq
A well-defined exponential of $d/du$ on step functions can be defined via
\beq
	\exp (c \frac{d}{du}) g(u)  
	:= \lim_{\e\rightarrow0} \exp (c \frac{d}{du}) g_\e(u)
	= g(u+c).
\eeq   
This is the definition we choose for $\frac{d}{du}\langle W_{\eta_u}\rangle$. 

The assumption is that (A1) can be combined with the regularization
(\ref{smearing}) such that $\frac{d}{du}\langle W_{\eta_u}\rangle = \langle
\frac{d}{du} W_{\eta_u} \rangle$.  For example, for $h(u) = \exp g(u)$,
which behaves under differentiation somewhat similar to $\langle
\frac{d}{du} W_{\eta_u}\rangle$, this scheme is consistent.

As a consequence of the above discussion one can now show that in fact
\beq
	\pupdown = \pn.
\label{takeapart}
\eeq 
While $\D \pupdown$ is not necessarily zero, we can define a new family of
loops that describes moving the two loops in $\pupdown$ apart (rather than
lifting half a leg in each of them as the original $\D$ does).  This is
obvious if we choose a family of loops in the plane of the tangent vectors,
since then $V=0$.  It is less obvious if we consider lifting one of the
loops out of the plane of the tangent vectors, but the generator $\D$ for
this operation actually annihilates $\pupdown$, as a calculation similar to
the one in section 2 shows. Exponentiation leads to (\ref{takeapart}), if
we also take into account that corners of a loop may be smoothed out, which
again can be shown by an analogous argument.

\section{Formal calculation of loop variations to all orders and
derivation of Witten's identity}

In section 3 we have given some detailed arguments that make the first
order calculation of section 2 more rigorous, in particular we have argued
that
\beq
	\D \pc = a \pc + b \pn
\label{dpc}
\eeq
for some finite coefficients $a$ and $b$. Furthermore, by the construction
of section 3.2 we have that
\beq
	e^{\D} \pc = \pp, \quad e^{-\D} \pc = \pp.
\label{epc}
\eeq
In order to calculate the left hand side of the equations in (\ref{epc}),
we have to know ${\D}^n\pc$, or $\D\pn$. Suppose that
\beq
	\D 
	\left(\begin{array}{c} \pc \\ \pn \end{array} \right)
	= 
	\left(\begin{array}{cc} a & b \\ c & d \end{array} \right)
	\left(\begin{array}{c} \pc \\ \pn \end{array} \right) .
\label{dmatrix}
\eeq
Then $e^{\D}\pc$ is known as a linear combination of $\pc$ and $\pn$, and
by eliminating $\pc$ in (\ref{epc}) we find that
\beq
	\a_+ \pp - \a_- \pmi = \a_\asymp \pn,
\label{generalp}
\eeq
where the coefficients are functions of $a$, $b$, $c$, and $d$.

The point is that once we know that $\D$ acts as the matrix transformation
(\ref{dmatrix}), the skein relation (\ref{generalp}) follows without
further assumptions about the path integral. In other words, a particular
type of skein relation follows already from a generic condition on how $\D$
acts.  

The reason that we present the derivation of the skein relation in such a
general manner is that it is not obvious how one should evaluate $\D\pn$.
The main problem is that $e^{\D}\pupdown$, if interpreted as the finite
deformation of the loop as in $e^{\D}\pc$, is not the expectation value of
Wilson loops. The result is 
$e^{\D}\pupdown \sim \langle \tr U_{\c} \tr U_{\c'} \rangle + ...$, 
where $\c$ and $\c'$ are open paths. In particular, the trace
of the parallel transport along an open path is not gauge invariant.

Let us therefore turn things around and ask whether there exist
matrices $\D$ at all such that we obtain Witten's result, i.e. that
\beq
	\a_+ = q^\fon, \quad \a_- = q^{-\fon}, \quad \a_\asymp = q-q^{-1}.
\label{coeffs}
\eeq
It is important to notice that if so, this is a non-trivial statement,
since as we will see below, arbitrary choices of $\D$ do not
generate all possible skein relations.

Instead of determining the matrix $\D$ that leads to the `correct' skein
relation directly from the explicit formulas that exist for the exponential
of a 2$\times$2 matrix, which are subject to various conditions on the
matrix, it is simpler and perhaps more intuitive to fix $\D$ by an order
by order analysis in an expansion in $1/k$ and then to check whether
exponentiation gives the right result to all orders.

The outcome of such an analysis is that for 
\beq
	q = \exp \frac{\pi i}{N+k},
\eeq
there does not exist a matrix $\D$ that leads to Witten's result. (The
adaptation of the coefficients fails at order $1/k^3$.) This has to be
expected, since each power of $\D$ produces an overall factor of $\pi i
/k$, and it is hard to see how these coefficients can combine to $\pi i /
(N+k)$, although a priori the possibility of a suitable `resummation'
cannot be excluded.

There does exist a matrix $\D$ that leads to Witten's result without the
shift in $k$, i.e. for
\beq
	q = \exp \frac{\pi i}{k}.
\label{q}
\eeq
It is a well-known problem of certain approaches to perturbative
Chern-Simons theory \cite{GuMaMi90,Aw90,Bar90} that they do not reproduce
the shift $k\mapsto N+k$, and our method seems to be of the same type. It
is sometimes argued that the shift is due to an independent argument about the
existence of the path integral --- this is consistent since $k$ appears only in
one place (in the definition of $q$) --- or that there is no need for the
shift \cite{Gu93}.

Let us sketch the calculation. We explicitly compute the matrix ${\D}^n$ to
find
\beq
	\pp = \sum_{n=0}^m \frac{1}{n!} {\D}^n \pc + O(\frac{1}{k^{m+1}}),
\eeq
from which we obtain the coefficients $\a_+$, $\a_-$, and $\a_\asymp$ in
(\ref{generalp}) up to order $m$. We try to choose $a$, $b$, $c$, and $d$
such that these coefficients agree order by order in
$1/k$ with (\ref{coeffs}) for  
\beq
	q^x = \sum_{n=0}^m \frac{1}{n!} (\frac{\pi i}{k} x)^n 
	+ O(\frac{1}{k^{m+1}}),
\eeq
for some exponent $x$. We find that 
\beqa
	\mbox{1. order} &\Rightarrow& a = - \frac{\pi i}{kN}, \quad
	b = \frac{\pi i}{k},
\\
	\mbox{2. order} &\Rightarrow& c = 0,
\\
	\mbox{3. order} &\Rightarrow& d = \frac{\pi i}{k} (\pm 1-\fon).
\eeqa
The two choices for $d$ turn out to be equivalent, so we pick the plus sign.
This fixes the available freedom. 

Since
\beq
	\exp \left(\begin{array}{cc} a & 1 \\ 0 & b \end{array} \right)  
	=
	\left(\begin{array}{cc} e^a & \frac{e^a-e^b}{a-b} 
	\\ 0 & e^b \end{array} \right)
\eeq
for $a\neq b$ (see below), we have that 
\beq
	e^{\D} \pc = q^{-\fon} \pc + q^{-\fon}(q-1) \pn ,
\label{edpc}
\eeq
and the corresponding equation for $e^{-\D}$ from $q\mapsto q^{-1}$. The
end result is the correct skein relation, (\ref{skeink}),
\beq
	q^\fon \pp - q^{-\fon} \pmi = (q-q^{-1}) \pn.
\eeq

To arrive at Witten's identity, we also have to derive the framing
dependence of $\psi_\c = \langle W_\c[A] \rangle$. The appropriate generator,
$\DD$, was already defined in section 3. By definition,
\beq
	e^{\DD} \psi_< = \ptp.
\eeq
Since $\DD \psi_<$ is proportional to $\psi_<$, exponentiation is trivial
and we obtain from (\ref{ddp}) that
\beqa
	\ptp &=& \a \psi_\wr, \quad \ptm = \a^{-1} \psi_\wr,
\\
	\a &=& q^{N - \fon},
\eeqa
which is the correct framing relation, (\ref{twistk}).

What we have to show in order to derive Witten's identity is that
\beq
	\psi_\c = K_\c,
\eeq
where $K_\c$ is defined by
(\ref{invk},\ref{normk},\ref{skeink},\ref{twistk}). We just established the
skein relation (\ref{skeink}) and the framing relation (\ref{twistk}).
That $\psi_\c$ is a regular isotopy invariant, (\ref{invk}), follows from
section 3.1, where we argued that $\psi_\c$ is an invariant of framed
loops. The normalization of $\psi_\c$ corresponds to a normalization of the
path integral and can be freely specified as in (\ref{normk}).

This concludes the formal derivation of Witten's identity. Linear order
considerations led us to assumptions (A1) and (A2). If we insist that the
exponentiation of $\D$ leads to Witten's identity in the particular scheme
that we introduced, then we are forced at third order to make the
assumption
\bi
\item[(A3)]
\beq
	\D \pupdown = \frac{\pi i}{k} (1 - \fon) \pupdown,
\eeq 
\ei
which we interpreted as an assumption about regularization. The simplicity
of the proposal lies in the fact that (A1)--(A3) reduce the
derivation of Witten's identity to essentially three functional
integrations with respect to $\eta^a$ and $A_a^i$ and the exponentiation of
a 2$\times$2 matrix.

\section{Discussion}

Let us discuss (A1)--(A3) and comment on two applications of our method.
\bi

\item
On (A1). \\
In section 3.2 we argued that to commute the limits of differentiation and
integration makes sense for $\frac{d}{du}\langle W_{\eta_u} \rangle$ if we
smooth out the step function in $u$. There are no conclusive arguments why
one should regulate this way, but the construction seems natural enough.

\item
On (A2). \\ 
For a generic definition of the measure $DA$ and the boundary
conditions, the assumption that in a partial integration there are no extra
terms is certainly wrong. It would seem that this issue cannot be discussed
without invoking the machinery of standard perturbation theory to
explicitly define a measure and a gauge fixing. However, very recently
Ashtekar and Lewandowski \cite{AsLe93} have constructed explicitly a
diffeomorphism and gauge invariant measure on $\overline{{\cal A}/SU(N)}$,
i.e.\ on a suitable completion of the space of connections modulo gauge
transformations involving distributions (see \cite{Bae93} for an extension).
This may come as a surprise and be of some importance for the path integral
formulation, since as generally believed there do not exist such measures
on ${\cal A}/SU(N)$ (which is neither Hausdorff nor locally-compact).

Although the Ashtekar-Lewandowski measure $(dA)_{AL}$ is not the correct
measure for our purposes, we consider the fact that such measures do exist as
encouragement that naive manipulations of the path integral may make sense
after all, in particular without gauge fixing. For example:
\bi

\item
Notice that a regulated version of 
\beq
	\hat T^a[\eta](s) = \tr U_\eta(s) \frac{\d}{\d A_a(\eta(s))}
\eeq
is a self-adjoint operator with respect to $(dA)_{AL}$
\cite{AsIs92,AsLe93}. $\hat T^a$ is the generator of diffeomorphisms (plus
an irrelevant gauge transformation) in the connection representation of
canonical Yang-Mills theory \cite{As91}. Let us define the Chern-Simons path
integral in terms of $(dA)_{AL}$ and assume that the integrand is
integrable. Wilson loops are integrable, but it is not clear that 
$
\exp(\frac{ik}{4\pi}S_{CS}) 
$ is integrable (which may make a regularization and renormalization
necessary). 
Then in the calculation of loop
variations in section 2 we have that
\beqa
	&&
	\int\!(dA)_{AL} \,
\exp(\frac{ik}{4\pi}S_{CS}) 
F^j_{ab} \tr U_\eta(s) T^j 
\nonumber
\\	
	&=&  
	- \frac{4\pi i}{k}
	\int\!(dA)_{AL}\, 
\e_{abc} \hat T^c [\eta](s)
\exp(\frac{ik}{4\pi}S_{CS}) 
\label{left}
\\
 	&=& 0,
\eeqa
since $\hat T^a$ is self-adjoint (and $\hat T^a\, 1 = 0$).  That the result
is zero is actually consistent with the definition of $(dA)_{AL}$, for
which the integral over a Wilson loop is zero if the loop is traversed an
odd number of times, so that also the left hand side of (\ref{left}) is
expected to be zero. Hence, in this scenario partial integration is
well-defined and trivial in the sense of (A1), although the result, while
diffeomorphism invariant, is not Witten's identity.

\item
Suppose that $DA$ is a diffeomorphism invariant measure such that the
Chern-Simons expectation value exists. Recall the discussion of section 3.1
where we argued for regulating the $\d(u)$ singularity. Suppose now that
we do not smear over $u$ and decide to deal with the derivative of the step
function directly. Then since
\beq
	\pp - \pc = \int_0^1\!du\, \frac{d}{du} \psi[\eta_u],
\eeq
we would conclude from (\ref{dp}) that
\beq
	\pp - \pc = \frac{\pi i}{k} (\pn - \fon \pc)
\eeq
is {\em exact}, i.e.\ satisfied without approximation in $1/k$.  
The obvious conclusion is that the necessary corrections arise from the 
partial integration. The term that we have neglected is
\beq
	- \frac{4\pi i}{k} \int_0^1\!du\int\!ds\,
	\e_{abc} v^a(s) \dot\eta_u^b(s) 
	\intA 
	\frac{\d}{\d A_c^j}
	\left(
	\exp(\frac{ik}{4\pi}S_{CS}) 
	\tr U_{\eta_u}(s) T^j 
	\right).
\eeq
If we assume that $DA = dA\, \mu(A)$, where $\d/\d A_c^j$ is self-adjoint with
respect to $dA$, then we can obtain an identity for the weight $\mu(A)$.
For example, if
\beq
	\e_{abc} \frac{\d}{\d A_c^j} \mu(A) = F^j_{ab} \mu(A),
\eeq
we obtain a series in $1/k$, which is the feature we are looking for, but
the result is wrong for two reasons. Witten's identity cannot be obtained
from $\pn - \fon \pc$ times an overall factor, and we need a factor $1/n!$
for an exponential. If such an iteration is successful, one should also
recover the shift.

\ei
Suppose we take the point of view that there exists a diffeomorphism
invariant measure similar to $(dA)_{AL}$, for which (A1) and (A2) can be
made rigorous by certain corrections. Then the curious fact remains that
the `mistakes' we make by assuming (A1) and (A2) can be compensated by
(A3).   

\item
On (A3). \\ The result that there is a matrix $\D$ as in (\ref{dmatrix}) that
leads to Witten's identity is a non-trivial feature of our method since
(\ref{dmatrix}) cannot accommodate arbitrary skein relations. Let us look
at a representative example for generic $\D$. For 2$\times$2 matrices $A =
\left(\begin{array}{cc} a & b \\ c & d \end{array} \right) $, it is simple
to compute $e^A$, since
\beq
A = C 
\left(\begin{array}{cc} \lambda_1 & 0 \\ 0  & \lambda_2 \end{array} \right)
C^{-1} \quad \Rightarrow \quad e^A = C
\left(\begin{array}{cc} e^\lambda_1 & 0 \\ 0  & e^\lambda_2 \end{array}
\right) C^{-1}, 
\eeq
if $C$ exists.
If $ad-bc\neq0$ and $(a-d)^2+4bc>0$, then a possible choice is
\beqa
	C &=& 
\left(
\begin{array}{cc} \lambda_1 - d & b \\ c  & \lambda_2-a \end{array}
\right) 
	, \quad\quad
	C^{-1} = \frac{1}{\det C} \left(
	\begin{array}{cc} \lambda_2 - a & -b \\ -c  & \lambda_1-d \end{array}
	\right) ,
\\
	\lambda_{1,2} &=& \fot ( a + d \pm \sqrt{(a-d)^2+4bc} \, ).
\eeqa
Hence,
\beqa
	\a_+ &=& ( a_1 e^{-\l_1} - a_2 e^{-\l_2}) / {\det C} ,
\label{aa1}
\\
	\a_- &=& ( a_1 e^{\l_1} - a_2 e^{\l_2}) / {\det C} ,
\\
	\a_\asymp &=& b_1 (a_1-a_2) 
	( e^{\l_2-\l_1} - e^{\l_1-\l_2}) / (\det C)^2 ,
\label{aa2}
\eeqa
where $a_1 = (\l_1-d)(\l_2-a)$, $a_2 = bc$, and $b_1=b(\l_1-d)$.

Notice that for such $A$, one does not in general obtain a skein relation
of the general type \cite{CoGuMaMi90}
\beq
	\b \pp - \b^{-1} \pmi = z \pn,
\label{gp2}
\eeq
for some complex coefficients $\b$ and $z$, since in general $\a_+ \neq
\a_-^{-1}$. At the same time, the coefficients cannot be made arbitrary
(even if $A$ is more special).

Let us address the question whether there is a natural interpretation for
the particular choice of $\D\pc$ made in (A3). There are, in fact, two
natural suggestions for $\D\pc$, which however lead to the wrong result.
One proposal that unfortunately fails is to just repeat the calculation of
$\D \pc$ of section 2 for $\D (\D\pc)$. The result is 
\beq
	(\D_1)^2 \pc := \D (\D \pc) \sim \fot \langle 
	\tr U_\c T^iT^j U_{\c'}T^iT^j  +
	\tr U_\c T^iT^j U_{\c'}T^jT^i 
	\rangle .
\eeq
From this follows (or directly from $\D\pupdown$ with due care about the
corners) that
\beq
	\D_1 
	\left(\begin{array}{c} \pc \\ \pn \end{array} \right)
	= 
	\frac{\pi i}{k}
	\left(\begin{array}{cc} - \fon & 1 \\ \fot & 
	\frac{N}{2} - \fon \end{array} \right)
	\left(\begin{array}{c} \pc \\ \pn \end{array} \right) .
\eeq
This matrix is of the general type that we just analyzed, but the resulting
skein relation is not even that of a knot polynomial of the type
(\ref{gp2}). There is no obvious reason why one should subtract out  
the terms leading to (A3).

Another idea could be to postulate that
\beq
	\D_2 \pupdown = 0.
\eeq
This amounts to mixing the limit in which $\pupdown = \pn$ with the
limit of differentiation, the idea being that $\D_2 \pn$ is zero because
the loops are separated. Previously we smoothed out the corners always
after all derivatives are taken. The resulting skein relation is
\beq
	q^\fon \pp - q^{-\fon} \pmi = N (q^\fon - q^{-\fon}) \pn, 
\label{wrong}
\eeq
which does fit the definition of a knot polynomial, although not Witten's
identity. The resulting framing relations are not of the type 
$\psi_{\propto\pm} = \a^{\pm1} \psi$. 
Notice that $c=0$ implies skein relations of the type (\ref{gp2}), while the
choice for $d$ given by (A3) is distinguished since then $\a_\asymp$ is
independent of $N$.

If one decides to introduce a relative factor unequal to one in
$\pupdown\sim\pn$ that adjusts (\ref{wrong}) to the correct result
(\ref{skeink}), then one runs into inconsistent framing relations. A simple
check of the self-consistency of the skein and framing relations are given
by the Mandelstam identities. E.g.\ for $SU(2)$ matrices $A$ and $B$, $\tr
A \tr B = \tr AB + \tr AB^{-1}$, which leads to some simple consequences
that are correct for $\D$ but incorrect for $\D_2$ with the rescaling.

\item
Abelian Chern-Simons theory. \\ 
Assumptions (A1) and (A2) and the regularization may be explicitly testable
in abelian Chern-Simons theory. The relevant equations in this case are 
\beqa
	\pc &=& \pupdown,
\\
	\D \pc &=& \frac{2\pi i}{k} \pc,
\\
	\psi_{\pm} &=& q^{\pm 2} \pn.
\eeqa
Notice that the last equation implies the well-known result that all the
information of $\psi_\c$ is in $w(\c)$. In particular, it would be
interesting to find out whether there exists a direct relation to
perturbation theory (e.g.\ \cite{CoMa90}).

\item
Skein relations for intersections. \\
As mentioned in the introduction, the extension of knot invariants to
intersecting loops is of interest in quantum gravity. From (\ref{edpc}) we
obtain immediately the skein and `framing' relations for intersections,
\beqa
	&& q^\fon (1 - q^{-1}) \pp - q^{-\fon} (1-q) \pmi = (q-q^{-1}) \pc,
\label{crossfactor}
\\
	&& \psi_\propto = \frac{ q^N (1-q^{-1}) - q^{-N} (1-q)}{q-q^{-1}} 
	\psi_\wr.
\label{twistfactor}
\eeqa
Independent of Witten's identity, a two variable version of the Jones
polynomial for intersecting loops, $F_\c(q,a)$, can be constructed from the
braid group with intersections \cite{Ga92,BrGaPu92b}. Skein relation
(\ref{crossfactor}) corresponds for $N=2$ precisely to the choice
$a=1-q^{-1}$ in $F_\c(q,a)$. The functional $\pc$ is therefore well-defined
in its own right, and it is not just a meaningless variable that is
introduced for purely technical reasons and that has to be eliminated from
the relation between $\pp$, $\pmi$, $\pn$, and $\pc$.

What we also would like to point out is that these relations are
non-trivial extensions of the linear order results, and have to be
contrasted with the defining relations of the Vassiliev invariants. (Their
relationship is discussed in \cite{Bae92,Ka93}.) To linear order we have
$\pc = \fot(\pp + \pmi) + O(\frac{1}{k^2})$ and $\psi_\propto = \psi_\wr +
O(\frac{1}{k^2})$. On the other hand, the characteristic relation for
intersections of the Vassiliev invariants is $V_\times = V_+ - V_-$
(no $k$ dependence).

\ei

In conclusion, loop deformations in $\langle W_\c[A] \rangle$ allow us to
derive skein relations for $\langle W_\c[A] \rangle$ that are to linear
order in $1/k$ that of $K_\c$.  The necessary assumptions (A1) and (A2) are
supported by results in perturbation theory, and by the fact that we obtain
the correct answer. A novel aspect of the construction is that we define
a generator of loop deformations that can be exponentiated.  The final
result depends on assumption (A3) for $\D
\pupdown$, which is perhaps less well-founded than (A1) and (A2), but 
gives the correct answer to all orders. What we would like to find is some
further evidence for (A3), say from an explicit regularization.

The heuristic level of our discussion does not allow us to decide whether a
rigorous proof of Witten's identity can be constructed from (A1)-(A3).  On
the other hand, we find it remarkable that all the uncertainty about the
measure $DA$ can be condensed into such a simple set of assumptions.

\subsection*{Acknowledgements}
It is a pleasure to thank P. H\'aji{\v c}ek, A. Higuchi, M.
Niedermaier, and J. Pullin for interesting discussions.

\newcommand{\bib}[1]{\bibitem[#1]{#1}}
\newcommand{\cqg}[1]{{\em Class.\ Quan.\ Grav.\ }{\bf #1}}
\newcommand{\grg}[1]{{\em Gen.\ Rel.\ Grav.\ }{\bf #1}}
\newcommand{\np}[1]{{\em Nucl.\ Phys.\ }{\bf #1}}
\newcommand{\pr}[1]{{\em Phys.\ Rev.\ }{\bf #1}}
\newcommand{\prl}[1]{{\em Phys.\ Rev.\ Lett.\ }{\bf #1}}
\newcommand{\pl}[1]{{\em Phys.\ Lett.\ }{\bf #1}}
\newcommand{\jmp}[1]{{\em J. Math.\ Phys.\ }{\bf #1}}
\newcommand{\jgp}[1]{{\em J. Geom.\ Phys.\ }{\bf #1}}
\newcommand{\cmp}[1]{{\em Commun.\ Math.\ Phys.\ }{\bf #1}}
\newcommand{\mpl}[1]{{\em Mod.\ Phys.\ Lett.\ }{\bf #1}}
\newcommand{\ijmp}[1]{{\em Int.\ J. Mod.\ Phys.\ }{\bf #1}}
\newcommand{\apny}[1]{{\em Ann.\ Phys.\ (N.Y.) }{\bf #1}}

\end{document}